\documentclass[aps,prb,twocolumn,preprintnumbers,amsmath,amssymb,floatfix]{revtex4}
\usepackage[dvips]{graphicx}
\usepackage{mathptmx}
\usepackage{verbatim}

\begin{document}
\title{Valley-Hall Kink and Edge States in Multilayer Graphene}
\author{Jeil Jung}
\affiliation{Department of Physics, University of Texas at Austin, USA}
\author{Fan Zhang}
\affiliation{Department of Physics, University of Texas at Austin, USA}
\author{Zhenhua Qiao}
\affiliation{Department of Physics, University of Texas at Austin, USA}
\author{Allan H. MacDonald}
\affiliation{Department of Physics, University of Texas at Austin, USA}
\date{\today{}}

\begin{abstract}
We report on a theoretical study of one-dimensional (1D) states
localized at few-layer graphene system ribbon edges,
and at interfaces between few-layer graphene systems with
different valley Hall conductivities.
These 1D states are topologically protected when
valley mixing is neglected.
We address the influence on their properties of stacking arrangement,
interface structure, and external electric field perpendicular to the
layers.
We find that 1D states are generally absent at multilayer
ribbon armchair direction edges, but present
irrespective of crystallographic orientation at any internal valley-Hall interface
of an ABC stacked multilayer. 
\end{abstract}



\maketitle

\section{Introduction}

Metallic surface states in a system with an insulating bulk are
often related to topological order.\cite{edgebulk} An important
example is provided by the quantum Hall effect of two-dimensional
(2D) systems in presence of an external magnetic field, in which
two-dimensional (1D) edge states \cite{halperin} accompany integer
valued Chern indices\cite{tknn} of bulk 2D bands. The recent
identification of topological insulator materials\cite{tireferences}
has provided a new example, one which does not rely on external
magnetic fields. In topological insulators strong spin-orbit
interactions yield bulk bands that have non-trivial values of a
Z$_{2}$ topological index and support topologically protected
surface states. The present study explores a 2D bulk example in
which the relevant topological index is, as in the quantum Hall
case, an integer-valued Chern index even though no magnetic field is
present.

Our work is motivated by the suggestion of Martin, Blanter, and Morpurgo
\cite{morpurgo} that 1D states can be
induced in bilayer graphene by changing the sign of an inter-layer
electric field.  (Electric field below will always refert to a field directed
between the layers of a few-layer graphene system.)  These states have a formal structure
similar to that of the zero modes
that appear in the A phase of $^3$He thin films at domain walls between regions with opposite
spontaneous orbital moments.\cite{volovik}
The 1D states in Ref.[\onlinecite{morpurgo}]
can be understood as being a consequence of separate Chern indices of
opposite sign associated with
the $K$ and $K'$ Dirac points of bilayer graphene.  The property that these
Chern numbers are implied by the momentum space Berry curvature
of bilayer graphene's bands\cite{xiaodi} when the electric field is non-zero
is referred to as the valley Hall effect.
The valley Chern numbers emerge
when the electric field breaks inversion symmetry to open a gap in the 2D bulk bilayer
electronic structure.  However, because the two valleys share the Brillouin-zone of a bilayer
graphene crystal, separate Chern numbers are never precisely defined and
are strictly speaking an artifact of the commonly employed continuum $\vec{k} \cdot \vec{p}$
electronic structure model.  Correspondingly, the 1D states that are the subject of
Ref.[\onlinecite{morpurgo}] are not guaranteed to be present at all energies and
are not topologically protected against perturbations that couple different valleys.
The goal of the present study is to assess the degree to which these caveats
are practically important.

1D states of the type we consider were first found in numerical
studies of zigzag bilayer ribbons \cite{castro}, and later
recognized as valley Hall edge states by Morpurgo {\em et al}.
\cite{morpurgo2} They have also been previously studied
theoretically in single-layer graphene samples with an imposed
staggered potential,\cite{semenoff,yao} {\em i.e.} a potential that
has opposite signs on the graphene honeycomb's A and B sublattices.
A more detailed analysis of the correspondence between bulk and edge
in bilayer graphene has been carried out recently, concluding that
the existence of 1D states at an edge depends on its
morphology.\cite{morpurgo3} In the present paper we consider both
bilayer graphene and other few layer graphene systems in which
electric fields open up an energy gap and yield a valley Hall
effect.  Continuum model considerations suggest that 1D states
should be present at the edge of graphene ribbons with a valley Hall
effect and at interfaces between systems with different valley Hall
effect quantum numbers.  We will refer to the former type of 1D
state as an edge state and to the latter as a kink state.  In the
case of bilayer graphene, for example, an interface that supports
kink states is easily produced by changing the sign of the electric
field along a line inside the material.
To create the corresponding 1D kink states in monolayer graphene, it
would be necessary to change the sign of the staggered potential.

In this paper we use $\pi$-band tight-binding models
to assess the influence of stacking orders and edge geometries
on edge and kink state properties.
As in the bilayer case, we find that the valley Hall edge states in
multilayers do not survive for armchair edge
terminations. However, kink states
are clearly present for both zigzag and armchair crystallographic
orientations of an internal interface along which the valley Hall
quantum number changes.
Section II contains the main results of our work.
We start discussing the valley Hall conductivity in terms of the 
low energy continuum model of ABC stacked $N$-layer graphene.
We conclude that we can normally expect $N$ 1D kink state branches 
per valley localized along electric field sign-change lines.
Then we use $\pi$-orbital tight-binding calculations on multilayer ribbons to
test the continuum model, presenting results for the energy bands of valley 
Hall edge and kink states for a number of different cases.
Finally in Sec. III we close with a brief summary and a discussion of our findings.


\section{Valley-Hall edge and kink states in multilayer graphene}
The electronic structure results for multilayer graphene ribbons presented here were
obtained using a  $\pi$-orbital tight-binding model Hamiltonian with nearest-neighbor
hopping and a lattice position dependent external potential $U_{i}$
\begin{eqnarray}
H &=&  - \sum_{<i,j>}   \gamma_{i,j} \,\,  c_i^{\dag} c_j      +  \sum_{i} U_{i} \, \, c_{i}^{\dag} c_{i}.
\end{eqnarray}
The hopping amplitude $\gamma_{i,j}$ we used is equal to $ t = 2.6 \,\, eV$ for in-plane hopping
and $t_{\perp} = 0.34 \,\, eV$ for out of plane hopping.
$c^{\dag}_i$ and $c_{j}$ are creation and destruction operators at $i^{th}$ and $j^{th}$ lattice sites.
This model can yield 1D edge state or kink state branches at energies 
inside bulk gaps which can be explained qualitatively in terms 
of the model's bulk valley Hall effect.   We start by discussing the valley Hall properties of multilayer systems
using a low energy model Hamiltonian, before returning to
tight-binding model calculations carried out for ABA and ABC stacked bilayers
trilayers, and tetralayer graphene ribbons.  We consider a variety of 
different examples that capture some essential features of stacked 
multilayer electronic structure.

\subsection{Valley Hall effect and associated 1D states in chirally stacked $N$-layer systems}
The low energy Hamiltonian in ABC stacked (rhombohedral) $N$-layer
graphene is useful as a simplified route for gaining insight of the
system despite the simplifying assumptions. In particular, the
valley Hall properties of ABC stacked (rhombohedral) $N$-layer
graphene can be derived from the low energy band structure in which
Bloch states are localized mainly on top and bottom layers. Because
the phase difference between top and bottom layer wavefunction
components varies more rapidly with momentum direction (measured
from the $K$ and $K'$ Dirac points) in larger $N$ systems, the
valley Chern numbers increase with $N$. \cite{hongki,trilayer}
To be more precise the wave functions of the states closest to
the Fermi level reside mostly on the top and bottom layer
lattice sites without a vertical neighbor, and the valley Chern number in the
presence of an electric field is equal to $N/2$ except possibly at
very low carrier densities and weak electric fields where weak band
structure features can play a role.
(Note that for odd $N$ this quantity is not an integer; we nevertheless refer to
the values as Chern indices for convenience.)
The sites with vertical neighbors have more weight in higher energy bands.

Strictly speaking, Chern numbers should be calculated by
integrating Berry curvatures over the whole Brillouin zone,
including contributions from near both $K$ and $K'$ points.
This total Chern number always vanishes, as it must 
when time-reversal symmetry is not violated.
In the continuum model approximation, however, we can speak of valley resolved
contributions to Chern numbers because the valley indices are
clearly distinguished.  In this picture, familiar quantum Hall considerations 
suggest the presence of 1D channels at edges and along lines where the 
valley Hall conductivity changes.  The microscopic tight-binding model
calculations we perform are intended to test the degree to which these 
considerations are reliable. 

\begin{figure}[htbp]
\begin{center}
\includegraphics[width=8.4cm,angle=0]{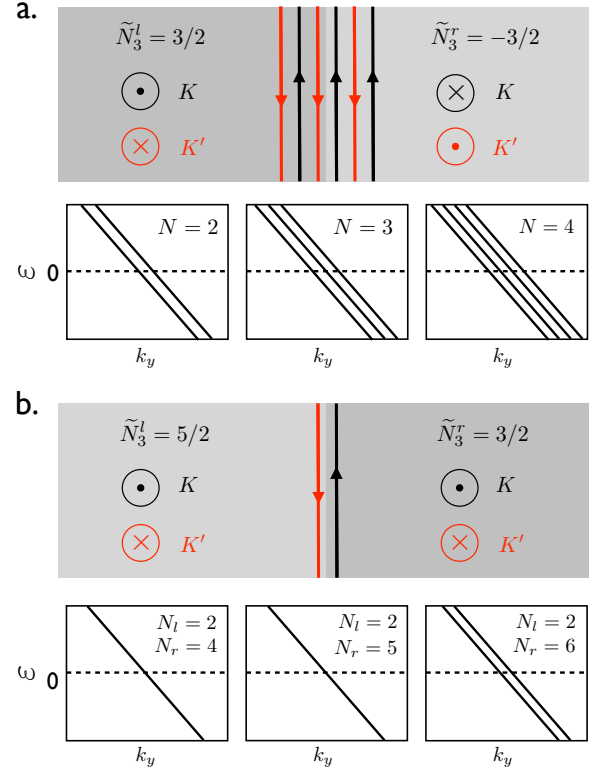}
\caption{
Schematic illustration of the relationship between valley Hall effects and 1D conduction channels
at interfaces expected on the basis of continuum model considerations.
The number of 1D modes per valley is an integer $n$ evaluated from differences
between valley Chern numbers $n = \left|  \widetilde{N}^{l}_{3} - \widetilde{N}^{r}_{3}  \right| $.
When the layer number $N$ is odd, the integral of the Berry curvatures
is half-odd-integer, a property that is related to the half-quantized Hall effect of Dirac systems.
We find that the number of 1D channels is then usually reduced to
the integer part of the integrated Berry curvature difference.
The sense of the arrows directed perpendicular to the page
indicates the sign of the Chern number associated with each valley.
{\em Upper panel:}
Illustration of ABC trilayer graphene with opposite electric field signs in left and right regions
with three 1D modes per valley.
{\em Lower panel:}
Junction formed between pentalayer and trilayer regions under a uniform bias potential.
In this case the valley Hall conductivities have the same sign but different magnitude
on opposite sides of the interface and the
number of 1D channels is expected to be proportional to the difference of
the individual Berry curvature integrals.
The continuum model picture illustrated here can be invalidated by
atomic scale physics at the interface, particularly when the number of layers on
opposite sides of the interface is different.
}
\end{center}
\label{schematic}
\end{figure}
Our working assumption in this section is that the following effective
continuum model\cite{hongki}  captures essential properties of
the full Hamiltonian:
\begin{eqnarray}
H_{N} &=& - t_{\perp} \left(
\begin{array}{cc}
0              &   \nu^{\dag N} \\
\nu^{N}    &     0
\end{array}
\right) +
\frac{1}{2}\left(
\begin{array}{cc}
\Delta               &   0   \\
0    &  - \Delta
\end{array}
\right)
\end{eqnarray}
where $-t_{\perp}$ is the interlayer hopping parameter,
$\nu = \upsilon_F \pi /t_{\perp}$,  $\upsilon_F \sim c/300$ is the Fermi velocity of
single layer graphene, $\pi = \hbar \left( k_x +  i k_y \tau_z \right)$ where
$\tau_z = \pm 1$ labels the $K$ and $K'$ valleys, $(k_x,k_y)$ is crystal momentum
measured from a Dirac point,
and $\Delta$ is the potential difference between layers produced by the electric field.
This Hamiltonian leads to momentum-space Berry curvatures that are
sharply peaked near $(k_x,k_y)=0$.  The valley Chern numbers are
obtained\cite{xiaodi,xiaofan,morpurgo,volovik} by integrating the Berry curvature over 2D momenta $(k_x,k_y)$
continued to $\infty$ to obtain ${C}_{\tau_z} = N  \tau_z  {\rm sgn}(\Delta) / 2$.
The valley Chern numbers are sometimes referred to as topological charges,
and denoted by $ \widetilde{N}_3$ 
as a reminder that the topology indices of gapped 2D systems 
may be thought of as a dimensional reduction of topological charge $N_3$ 
at gapless 3D Fermi points. \cite{volovik}
The contribution to the Hall conductivity from a particular valley
is  $\sigma^{\tau_z}_{xy} = \widetilde{N}_3  e^2 / h$. In a
continuum model, the number of one-dimensional (1D) channels per
valley at an interface between two bulk regions with different
valley Hall conductivities $\sigma^{\tau_z}_{xy} = \widetilde{N}_3
\, e^2 / h$ is equal to the difference between their valley Chern
numbers.\cite{volovik,morpurgo,yao} For ballistic transport each
channel contributes $e^2/h$ to the two-probe conductance. When the
sign of the electric field is reversed at an interface, the Chern
number difference is $ 2 \times (N/2) = N$, equal to the layer
number
as illustrated schematically in Fig.~1. 

The generalization of the notion of a valley Hall conductivities from 
one and two layer systems to general $N$ layer systems
can also be made using the explicit equations for the 1D interface states by 
following a procedure similar to that outlined in Ref. [\onlinecite{morpurgo}]:
\begin{eqnarray}
ÊÊ- V(x) u + K_{N} \left( \partial_x +  k_y \right)^N \upsilon    =  \varepsilon u  \nonumber    \\
ÊÊ  K_{N} \left( \partial_x  -  k_y \right)^N u + \,V(x) \upsilon  =  \varepsilon \upsilon.
\label{eq:continuum}
\end{eqnarray}
Here $K_N = -t_{\perp} \left( -i \,\upsilon_F \hbar / t_{\perp} \right)^{N}$,
$N$ is the number of layers in the system,
and $V(x)$ is a general position-dependent function which specifies
the difference between top and bottom layer potentials.
In the following subsections, we will present tight-binding
calculations for multilayers for a variety of different external potential
profiles
and discuss the validity of the qualitative picture summarized in Fig. 1.

\subsection{Multilayer ribbons under a uniform electric field}
The simplest example of valley Hall edge states are those that
appear in ABC stacked multilayers under a uniform external electric
field. Tight-binding band structure of bilayer graphene ribbons by
Castro {\em et al.}\cite{castro,xiaofan} has demonstrated the
presence of metallic edge states which cross the Fermi level in
neutral zigzag terminated bilayer ribbons and in chirally stacked
multilayer zigzag ribbons. The number of valley Hall edge states
branches in each propagation direction is equal to
$[N/2]$\cite{xiaofan}. Changing either edge termination or stacking
sequence introduces qualitative changes in the ribbon band structure
as we now describe.

In Fig. 2,  we plot band structures for ABC and ABA zigzag and armchair terminated
trilayer and tetralayer ribbons.
\begin{figure}[htbp]
\label{vhalledges}
\begin{center}
\includegraphics[width=6.6cm,angle=90]{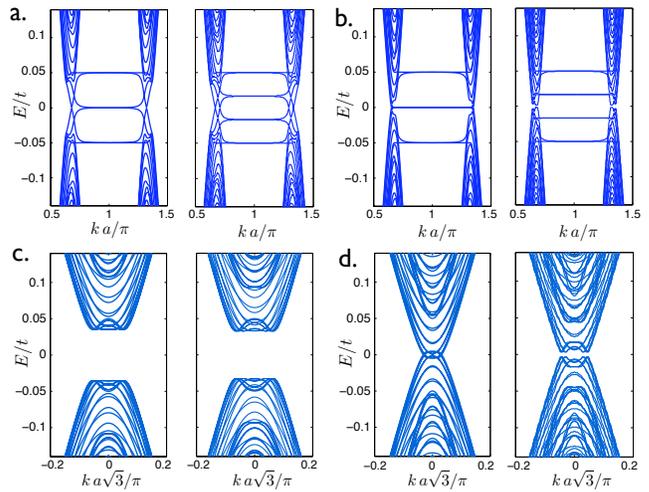}
\label{multilayer_zigzag}
\caption{
Band structures of ABC and ABA stacked zigzag and armchair terminated
trilayer and tetralayer ribbons in the presence of a uniform electric field.
(a) ABC stacked layers with zigzag edges; (b) ABA stacked layers with zigzag edges;
(c) ABC stacked layers with armchair edges and (d) ABA stacked layers with armchair edges.
In each subfigure, the left and right panels represent trilayer and tetralayer ribbons, respectively.
For ABC stacking the electric field opens a bulk band gap containing edge states
in the zigzag case but not in the armchair case.  The number of edge state branches
is independent of the ribbon width, whereas the number of bulk state branches is
proportional to ribbon width.
In the case of ABA stacking, ribbons are metallic in 
the trilayer whereas a small gap opens in the tetralayer geometry.
}
\end{center}
\end{figure}
In all these calculations we have maintained the same interlayer
potential difference $\Delta/t = 0.1$ between top and bottom layers
and have taken the electric field to be uniform. In the tetralayer
case the smaller potential difference $\Delta' = \Delta/3$ 
between the inner layers to facilitate distinguishing their edge state bands
associated with respect to the outer ones. In multilayer zigzag ribbons with
ABC stacking, the uniform electric field generates a band gap in the
sample bulk\cite{trilayer} with edge states in the gaps.
In agreement with previous analysis \cite{xiaofan}, we find $[N/2]$
valley Hall edge states in an $N$-layer ribbon in each valley that
propagate in opposite directions and are localized at opposite edges. 
Hence, for zigzag edges the number of 1D states at the physical boundaries 
of the ribbons with vacuum can also be discussed in terms of valley 
Chern number differences between the bulk region and 
the vacuum $\widetilde{N}_3^{vac} = 0$.
In the armchair termination both $K$ and $K'$ valleys appear at the
same projected 1D momentum and the Hall edge states are annihilated.
This observation is helpful in distinguishing edge and kink states
in ribbons with internal electric field sign changes.

For biased ABA multilayer zigzag ribbons the system also has
conducting edge states, but their pattern is more complex than in
the ABC case. (Bulk and edge ribbon bands can be distinguished by
their dependence on ribbon width.) In ABA stacked trilayer ribbons
an external bias increases the number of bulk channels, while it
opens a bulk gap in the ABC case.~\cite{trilayer,McCann,koshino}

\subsection{ABC stacked zigzag and armchair multilayer ribbons with kink states}
We have just seen that a biased bilayer and ABC trilayer graphene
ribbons are gapped in the bulk but have metallic valley Hall edge
states crossing the Fermi level when they have a zigzag edge
termination. Now we consider ribbons with a electric field sign
change at the ribbon center that is expected to produce $N$ 1D
channels in each valley. In these ribbons, the regions near the
ribbon border will show valley Hall edge states similar to those
present in a bilayer with a uniform electric field. 
In Fig. 3 we plot the band structures of bilayer, trilayer and tetralayer
ribbons with zigzag edge terminations subject to a step-like
interlayer potential discussed earlier.

In zigzag terminated ribbon geometries the $K$ and $K'$ momentum
projections appear at the two valley points located at 
$k = 2\pi/3a, 4\pi/3a$ and can therefore easily be distinguished.\cite{nakada} 
In agreement with the continuum model analysis presented in Sec. II, we
see $N$ bands of confined kink states with a well defined
propagation direction for each valley $K$ or $K'$ crossing the Fermi level. 
Each valley has doubly degenerate additional metallic edge state branches, 
with a velocity opposite to that of the confined states. 
In the uniformly biased case, 
edge states in a given valley that are localized on opposite edges 
have opposite propagation directions,
whereas they propagate in the same direction when a kink is present. 
For ribbons with inversion symmetry at the ribbon center, 
the co-propagating edge state channels are degenerate.

As mentioned previously, the projection of the 2D bands of graphene
to obtain the ribbon band structure places $K$ and $K'$ valleys at
the same momentum in the armchair edge case. Therefore, unlike the
case of the zigzag ribbons, it is not possible to identify valley
labels from ribbon band structure plots. In the case of edge states,
these difference eliminates the 1D channels completely. As we see in
in Fig. 3 this is not the case for ribbon states which appear to be as
robust in armchair and zigzag directions. When the electric field
profile has sharp spatial variation, there is a barely visible gap
opening, similar to the one found in direction of graphene under a
staggering potential \cite{semenoff}. This gap size decreases
quickly when the potential variation at the domain becomes smoother.

\begin{figure}[htbp]
\label{zigzag}
\begin{center}
\includegraphics[width=5.2cm,angle=90]{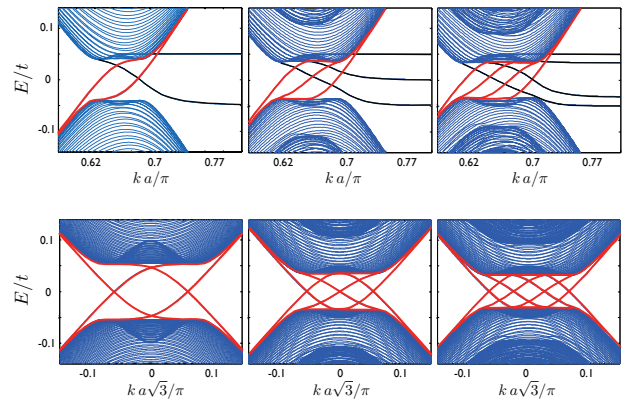}
\caption{(Color online)
{\em Upper Panel:} Band structure of bilayer, trilayer and tetralayer graphene
zigzag ribbons with an electric field sign change at the ribbon center.
We can clearly observe two, three and four 1D valley Hall kink states
with a common sign of velocity
located around the valley point.  The branches which correspond to
wavefunctions localized at the ribbon center are
plotted in red.  The other branches that cross the bulk gap are
doubly-degenerate edge states.
{\em Lower Panel:}
Band structure of bilayer, trilayer and tetralayer graphene armchair ribbons with
an electric field sign change at the ribbon center.
As in the zigzag case we can clearly identify two, three and four three
1D kink state branches.
In armchair edges we do not find 1D edge state channels.
}
\end{center}
\end{figure}

In the band structures of bilayer, trilayer and tetralayer graphene
armchair ribbons subject to a kink step bias around the ribbon
center, one can clearly identify two, three and four 1D states for
each propagation direction, corresponding to confined states at the
domain wall for wave vectors around $k \sim 0$ where both right and
left going states coexist with similar Bloch function wave vectors.
In armchair edges we do not find edge localized states crossing the
Fermi level as we had found for the zigzag terminated systems.

\section{Summary and conclusions}
External electric fields between layers give rise to gaps at the 
carrier neutrality point in bilayer and ABC multilayer graphene systems.
We have shown that 1D transport channels appear along lines 
where the sign of the inversion symmetry breaking
potential changes.  This finding generalizes results obtained previously for the 
bilayer \cite{morpurgo} and monolayer cases.
\cite{semenoff,yao} The number of these metallic 1D kink
state branches is proportional to the number of layers $N$ and 
can be related to the bulk valley Hall conductivity.\cite{xiaofan}. 
The states we have considered arise at
the boundary between two regions with opposite valley Hall
conductivity.  Because the Hall conductivity changes sign in opposite 
directions in the two valleys both produce 1D kink states and they 
propagate in opposite directions. 

Similar valley Hall effect considerations suggest that kink states should occur at boundaries between 
ribbons with different thicknesses.  
Our $\pi$-orbital tight-binding model calculations 
for ribbons with a trilayer/tetralayer boundary
find that the interface electronic structure depends on 
whether the stacking sequences is ABC or ABA. 
Bulk valley Hall effect values are also unreliable at the edge;
in particular we find that edge states are absent in
multilayer armchair terminated ribbons, as found earlier in the bilayer or
monolayer case.  This finding is perhaps expected since
the 1D momentum projection does not distinguish valleys in this case.
It is the robustness of the kink states at internal electric field sign changes
that is perhaps the surprise.  It is important to determine if it 
persists in the presence of disorder and turns in the
propagation path.

{\em Acknowledgments.}
The authors acknowledge helpful discussions with Qian Niu.
Financial support was received from Welch Foundation grant TBF1473,
NRI-SWAN, DOE grant Division of Materials Sciences and Engineering DE-FG03-02ER45958.

\end{document}